\documentclass{article}
\usepackage{amssymb}
\usepackage{amsmath}
\usepackage{amsfonts}
\usepackage{amsthm}
\usepackage[T1]{fontenc}
\usepackage{authblk}

\newtheorem{proposition}{Proposition}
\newtheorem{lemma}{Lemma}
\newtheorem{theorem}{Theorem}
\newtheorem{corollary}{Corollary}
\newtheorem{remark}{Remark}
\def\BOne{{\mathchoice {\rm 1\mskip-4mu l} {\rm 1\mskip-4mu l}
                          {\rm 1\mskip-4.5mu l} {\rm 1\mskip-5mu l}}}

\begin{document}

\title{Transverse K\"{a}hler-Ricci flow and deformations of the metric on the Sasaki space $T^{1,1}$}
\author[1,4]{Vladimir Slesar\thanks{vladimir.slesar@upb.ro}}
\author[2]{Mihai Visinescu\thanks{mvisin@theory.nipne.ro}}
\author[3,4]{Gabriel-Eduard
V\^ilcu\thanks{gvilcu@upg-ploiesti.ro}\thanks{gvilcu@gta.math.unibuc.ro}}
\affil[1]{University Politehnica of Bucharest, Department of Mathematical Methods and Models,
313 Splaiul Independentei, 060042 Bucharest, Romania}
\affil[2]{Department of Theoretical Physics,

National Institute for Physics and Nuclear Engineering,

Magurele, P.O.Box M.G.-6, Romania}
\affil[3]{Department of Cybernetics, Economic Informatics, Finance and Accountancy,

Petroleum-Gas University of Ploie\c sti,

Bd. Bucure\c sti, Nr. 39, Ploie\c sti 100680, Romania}
\affil[4]{Faculty of Mathematics and Computer Science,

Research Center in Geometry, Topology and Algebra,

University of Bucharest, Str. Academiei, Nr. 14, Sector 1,
Bucharest 060042, Romania}
\date{\today }
\maketitle

\begin{abstract}

In this paper we investigate the possibility to obtain locally new Sasaki-Einstein metrics on
the space $T^{1,1}$ considering a deformation of the standard metric tensor
field. We show that from the geometric point of view this deformation leaves transverse
and the leafwise metric intact, but changes the orthogonal complement of the Reeb vector
field using a particular basic function. In particular, the family of metric obtained using
this method can be regarded as solutions of the equation associated to the Sasaki-Ricci
flow on the underlying manifold.

\end{abstract}

\affil[1]{University Politehnica of Bucharest, Department of Mathematical Methods and Models,
 313 Splaiul Independentei, 060042 Bucharest, Romania}
\affil[2]{Department of Theoretical Physics,

National Institute for Physics and Nuclear Engineering,

Magurele, P.O.Box M.G.-6, Romania}
\affil[3]{Department of Cybernetics and Economic Informatics,

Petroleum-Gas University of Ploie\c sti,

Bd. Bucure\c sti, Nr. 39, Ploie\c sti 100680, Romania}
\affil[4]{Faculty of Mathematics and Computer Science,

Research Center in Geometry, Topology and Algebra,

University of Bucharest, Str. Academiei, Nr. 14, Sector 1,
Bucharest 060042, Romania}

\section{Introduction}

In the last time Sasakian geometries, as an odd-dimensional analogue of K\"{a}hler geometries,
have become of high interest in connection with some modern developments in mathematics,
see e.g. \cite{BG} and the references therein. In fact, Sasakian geometry has a twofold relation with
K\"{a}hlerian geometry: the Riemannian cone of a Sasaki manifold is K\"{a}hler, while the
1-dimensional foliation generated by the Reeb vector field is transversely K\"{a}hler.
Due to this fact, most concepts from K\"{a}hler geometry arise in a very natural way in Sasakian setting.
In theoretical physics, the interest in
Sasaki-Einstein geometry \cite{JS} has arisen due to its significant role in studies of
consistent string compactifications  and in the context of AdS/CFT correspondence.
In five dimensions new Sasaki-Einstein structures on $S^2 \times S^3$, denoted by $Y^{p,q}$,
have been constructed in \cite{GMSW} which contain the homogeneous space $T^{1,1}$
as a special case \cite{M-S}. The theory of contact structures is linked to many geometric
backgrounds as symplectic geometry, Riemannian and complex geometry, analysis and dynamics.

A well known method for generating Einstein metrics on Riemannian manifolds is the Ricci flow
originally introduced by Hamilton \cite{RH}. After being used by Perelman in proving the
famous Poincar\'{e} and Thurston conjectures, the Ricci flow became a fervent topic of
geometric analysis. It is also worth noting that using the Ricci flow approach,
Brendle and Schoen proved in 2009 the Differential Sphere Theorem \cite{BS}.
On the other hand, the complex analogue of Hamilton's Ricci flow, known as
K\"{a}hler-Ricci flow, was first used by Cao \cite{CAO} to give a parabolic proof of the
Calabi-Yau theorem. Recently the method was applied to Sasaki manifolds in \cite{S-W-Z} to
generate new Sasaki structures, the authors proving the well-posedness of the Sasaki-Ricci flow
and its long-time existence on compact manifolds. Recall that the Sasaki-Ricci flow consists in
deforming a Sasaki metric in such a way that the corresponding transverse K\"{a}hler metric
to be deformed along its Ricci curvature. Hence, the Sasaki-Ricci flow is simply the
transverse K\"{a}hler-Ricci flow, transforming this transverse K\"{a}hler structure in the
direction of the transverse Ricci curvature. Notice that Collins \cite{COL2016} obtained
two groups of conditions each of which guarantees the convergence of the normalized Sasaki-Ricci flow.
Moreover, Bedulli, He and Vezzoni consider a generalization to the K\"{a}hler-Ricci flow,
using this as a power tool to study K\"{a}hler foliations and obtain short-time existence and uniqueness
for solutions to the transverse K\"{a}hler-Ricci flow \cite{BHV}.

In this paper we investigate local deformations of Sasakian structures exploiting the transverse
structure of Sasakian manifolds. To be more specific, in the spirit of \cite{G-K-N}, using smooth functions
defined on local subsets of the five-dimensional Sasaki-Einstein space $T^{1,1}$, we consider local 
deformation of the Sasaki structure preserving the transverse metric. In this way, starting with the 
Sasaki-Ricci soliton $T^{1,1}$, we produce families of local Sasaki-Einstein metrics.

The paper is organized as follows. In Section 2 we review fundamentals on Sasaki geometry,
deformations of Sasaki metrics and Sasaki-Ricci flow. In Section 3 we deform the metric by keeping
the transverse and leafwise metric intact. For this purpose we consider perturbations of the contact
form using particular basic functions defined on open subsets. 
In order to apply the general scheme to the Sasaki-Einstein space
$T^{1,1}$, in Section 4 we introduce local charts and local coordinates, and construct the Sasaki analogue of the
K\"{a}hler potential. In Section 5 we produce families of Sasaki-Einstein metrics by deforming the orthogonal 
complement to the leaves. In Section 6 we show that these deformations of the $T^{1,1}$ metric can be
regarded as solutions of the equation of the Sasaki-Ricci flow. In the last Section we provide some closing remarks.

\section{Preliminaries}

In this section we review basic definitions and results concerning the geometry of Sasaki manifolds,
mainly based on \cite{BG,FOW,G-K-N}.

\subsection{Sasaki manifolds and Sasaki potential}

Let $(M,g)$ be a Riemannian manifold. Then the cone manifold $C(M)$ of $M$ is a Riemannian manifold
diffeomorphic to $(0,\infty)\times M$, equipped with the cone metric \[\bar{g}=dr^2+r^2g,\]
where $r$ is a coordinate on $(0,\infty)$. Recall that $M$ is said to be a Sasaki manifold if the
cone manifold $C(M)$ of $M$ has a K\"{a}hler cone structure $(J,\bar{g})$. Notice that any Sasaki
manifold $M$ is of odd dimension $2n+1$, where $n+1$ is the complex dimension of the K\"{a}hler
cone $C(M)$. It is clear that on $C(M)$ we have a vector field $\bar{\xi}$ and a 1-form $\bar{\eta}$
defined by:
\[
\bar{\xi}=Jr\frac{\partial}{\partial r},
\] 
and
\[
\bar{\eta}(\cdot)=\frac{1}{r^2}\bar{g}(\bar{\xi},\cdot),
\]
respectively. Moreover, the vector field $\bar{\xi}$ restricted to $M$ is a Reeb vector field
(let us note it by $\xi$) which induces a Reeb flow, while the 1-form $\bar{\eta}$ restricts to a
1-form $\eta$ on $M$. It is known that the basic functions are those functions which are invariant
under the flow generated by the Reeb vector field \cite{COLL}. Next, let us denote by $L_{\xi}$ the
line subbundle generated by $\xi$. We also consider the quotient transverse bundle $\nu$, defined as
$\nu=TM/L_\xi$. As in \cite{S-W-Z}, we consider the projection $p:TM \rightarrow \nu$. 
We get the exact sequence
\[
0 \rightarrow L_\xi \rightarrow TM \rightarrow \nu \rightarrow 0.
\]

Let now $\mathcal{D}=\mathrm{Ker}\eta$ be the contact subbundle in $TM$.
Then we have the following decomposition of the tangent bundle $TM$ of $M$:
\begin{equation} \label{splitting}
TM=\mathcal{D}\oplus L_{\xi},
\end{equation}
and we obtain a map $\sigma:\nu \rightarrow \mathcal{D}$ such that
\[
p\circ\sigma=\mathrm{Id}_{\nu}.
\]

We also derive that $M$ can be endowed with a contact structure $(\Phi,\xi,\eta)$, where
\[
\Phi_{|D}=J_{|\mathcal{D}},\ \Phi_{|L_{\xi}}=0,
\]
such that the 1-dimensional foliation $\mathcal{F}_\xi$ generated by the Reeb vector field $\xi$
is transversely K\"{a}hler. In this way, one gets a global 2-form $\Omega^T$ on $M$ coming from
the contact 1-form $\eta$, namely
\[
\Omega^T=\frac{1}{2}d\eta.
\]

We have that $(\mathcal{D},\Phi_{|D},d\eta)$ gives $M$ a transverse K\"{a}hler structure with
K\"{a}hler form $\Omega^T$ defined above and transverse metric $g^T$ given by
\[
g^T(X,Y)=d\eta(X,\Phi Y),
\]
and related to the Sasaki metric $g$ on $M$ by
\[
g=g^T+\eta\otimes\eta.
\]
Notice that  the transverse metric associated with  a Sasaki-Einstein space is Einstein.

It is known that a Sasaki structure on a Riemannian manifold $(M,g)$ can be also defined through
a unit length Killing vector field $\xi$ such that the Levi-Civita connection of the metric $g$
satisfies \cite{BG}
\[
(\nabla_X \Phi)Y=g(\xi,Y)X-g(X,Y)\xi,
\]
for all vector fields  $X,Y$ on $M$, where $\Phi X:=\nabla_X \xi$. Due to this reason, a Sasaki
structure on a smooth manifold $M$ can be denoted by a quadruple $(g,\Phi,\xi,\eta)$. However,
despite the notation, it is clear that the Sasakian structure is completely determined by the
knowledge of any pair of the following: $(g,\xi)$, $(g,\eta)$ or $(\Phi,\eta)$.

We recall next that a Riemannian manifold $(M, g)$ that satisfies the Einstein equation
\[
Ric_g=\lambda g,
\]
for a real constant $\lambda$ (called Einstein constant), where $Ric_g$ stands for the Ricci
tensor of the metric $g$,
is said to be an Einstein space or an Einstein manifold. Moreover, if the Einstein constant is zero,
then the Riemannian space $(M, g)$ is called a Ricci-flat manifold. A Sasaki manifold is said to be
a Sasaki-Einstein space if the cone manifold $C(M)$ of $M$ is K\"{a}hler Ricci-flat. It is clear that
a Sasaki-Einstein space is a Riemannian manifold that is both a Sasaki manifold and an Einstein space.

According to \cite{G-K-N},  every $(2n+1)$-dimensional Sasakian manifold is locally generated by a
free real-valued function $K$ of $2n$ variables defined on a local subset of $U$ of $M$, called 
the Sasaki potential, while every locally Sasaki-Einstein space of dimension $2n+1$ is generated 
by a locally K\"{a}hler-Einstein space of
dimension $2n$. Actually, if $U$ is a foliated chart on $M$ with
$U=I\times V$ (where $I\subset\mathbb{R}$ is an open interval and
$V\subset\mathbb{C}^n$), and $(x,z^1,\ldots,z^n)$ are the local holomorphic coordinates on
$U$ (with Reeb vector field  $\xi=\frac{\partial}{\partial x}$ and $z^1,\ldots,z^n$ are the
local holomorphic coordinates on $V$), then the Sasaki potential $K$ on $U$ is
chosen in such a way that $\xi(K)=0$ and
\[
\eta=dx+i\sum_{j=1}^n(K_{,j}{\rm d}z^j)-i\sum_{\bar{j}=1}^n(K_{,\bar{j}}
d\bar{z}^{j}),
\]
\[
d\eta=-2i\sum_{j,\bar{k}=1}^n K_{,j\bar{k}}
dz^{j}\wedge d\bar{z}^{k},
\]
\[g=\eta^2+2\sum_{j,\bar{k}=1}^nK_{,j\bar{k}}dz^j{\rm d}\bar{z}^{k},\]
\[\phi=-i\sum_{j=1}^n[(\partial_j-iK_{,j}\partial_x)\otimes dz^j]+i\sum_{\bar{j}=1}^n
(\partial_{\bar{j}}+iK_{,\bar{j}}\partial_x)\otimes d{\bar{z}}^{j}].\]

It is important to note that the Sasaki potential does not possess the property of uniqueness.
Therefore, some different Sasaki potentials may lead to the same Sasaki structure. For example,
the transformation
\[
K(z,\bar{z})\rightarrow K(z,\bar{z})+f(z^j)+\bar{f}(\bar{z}^j),\, x\rightarrow x+i\bar{f}(\bar{z}^{j})-if(z^j),
\]
where $f$ and $\bar{f}$ are arbitrary holomorphic and anti-holomorphic functions,
does not alter the Sasaki structure.

We recall that a  $r$-form $\alpha$ on $M$ is called \emph{basic} if
\[
\iota_\xi \alpha =0,\quad \mathcal{L}_\xi \, \alpha =0\,,
\]
where $\mathcal{L}_\xi$ is the Lie derivative with respect to the vector field $\xi$.
In particular a function $\varphi$ is basic if and only if $\xi(\varphi)=0$.
In the system of coordinates $(x,z^1,\ldots,z^n)$ given above, a basic $r$-form of type
$(p,q)\,,\, r=p+q$ has the form
\[
\alpha = \alpha_{i_1\cdots i_p \bar{j}_1 \cdots \bar{j}_q} dz^{i_1} \wedge
\cdots \wedge dz^{i_p} \wedge d\bar{z}^{j_1} \wedge \cdots \wedge d\bar{z}^{j_q}\,,
\]
where $\alpha_{i_1\cdots i_p \bar{j}_1 \cdots \bar{j}_q}$ does not depend on $x$. 
We denote by $\Omega_B$ the \emph{de\thinspace Rham complex of basic differential forms}. 
The restriction $d_{B}:=d_{\mid \Omega_B}$ is called the basic de\thinspace Rham operator. 
We also consider \emph{the canonical basic Dolbeault operators}
\[
\partial_B=\sum _{j=1}^{n}dz^j \frac{\partial}{\partial z^j}, \quad
\bar{\partial_B}=\sum _{j=1}^{n} d\bar{z}^{j} \frac{\partial}{\partial \bar{z}^{j}} .
\]
We get that $d_{B}=\partial_B+\bar{\partial}_B$; we also define $d_{B}^{c}=\frac{i}{2}(\bar{\partial}_B-\partial_B)$.

\subsection{Deformation of Sasaki metrics}

There are various ways to deform Sasakian structures. In the following, we will recall some of
them, with an emphasis on the transverse K\"{a}hler deformation, that is a special case of a
type II deformation introduced by Belgun \cite{Bel}, and studied also by Boyer, Galicki and Matzeu \cite{BGM}.

Let us denote by $\mathcal{S}(\xi)$ the space of all Sasaki structures on $M$ having a fixed
Reeb vector field $\xi$. For such a fixed $\xi$, we consider a settled transverse complex
structure on $\mathcal{F}_\xi$. We denote by $\mathcal{S}(\xi,\bar{J})$ the subset of
$\mathcal{S}(\xi)$ consisting in all Sasaki structures $(g,\Phi,\xi,\eta)\in\mathcal{S}(\xi)$
with the same transverse
holomorphic structure $\bar{J}$. Then we have the following result.

\begin{lemma}\label{Lema1} {\rm \cite{BG}}
The space ${\mathcal S}(\xi, \bar{J})$ of all Sasakian structures with Reeb
vector field $\xi$ and transverse holomorphic structure $\bar{J}$ is an affine
space modeled on $(C_B^{\infty}(M)/{\mathbb R})\times
(C_B^{\infty}(M)/{\mathbb R})\times H^{1}(M,{\mathbb Z})$. Indeed, if
$(\xi, \eta, \Phi, g)$ is a given Sasakian structure in
${\mathcal S}(\xi, \bar{J})$, any other Sasakian structure
$(\xi, \tilde{\eta},\tilde{\Phi}, \tilde{g})$ in ${\mathcal S}(\xi, \bar{J})$ is determined by
real valued basic functions $\varphi$ and $\psi$ and integral closed $1$-form $\alpha$,
such that
$$\begin{array}{rcl}
\tilde{\eta} & = & \eta+ d^c \varphi + d\psi +i(\alpha) \, , \\
\tilde{\Phi} & = & \Phi - (\xi \otimes (\tilde{\eta}-\eta)) \circ \Phi \, ,\\
\tilde{g} & =  & d\tilde{\eta}\circ (\BOne \otimes \tilde{\Phi} ) +
\tilde{\eta}\otimes \tilde{\eta} \, ,
\end{array}
$$
where $d^c=\frac{i}{2}(\bar{\partial} - \partial)$, and $i:H^{1}(M,{\mathbb Z})\mapsto
H^{1}(M,{\mathbb R})=H^{1}({\mathfrak F}_{\xi})$ is the homomorphism induced by
inclusion.
In particular, $d\tilde{\eta}=d\eta + i\partial\bar{\partial} \varphi$.
\end{lemma}

Notice that it is very useful and natural to consider the above type of deformation in terms
of basic forms on the Sasaki manifold \cite{FOW}.

A second deformation of interest in Sasaki geometry is given by the deformation of the
transverse complex structure $\bar{J}$ on ${\mathfrak F}_{\xi}$ \cite{NOZ}. A special interest
is granted to those deformations of $({\mathfrak F}_{\xi},\bar{J})$ that
preserve the smooth foliation structure \cite{VCOV}. Though any small deformation of the complex
structure of a compact K\"{a}hler manifold admits compatible K\"{a}hler metrics, it is known that
there exists a Sasakian manifold whose Reeb flow can be deformed so that it does not admit a
compatible Sasakian metric. In \cite{NOZ}, the author obtains an obstruction to the existence of
compatible Sasakian metrics given by the (0,2) component of the basic Euler class of flows.
More precisely, he proved the following result.

\begin{theorem} {\rm \cite{NOZ}} Let $(\mathcal{F}_\xi,\bar{J}_t),\ t\in\mathcal{B}$ be a
deformation of the Reeb foliation of $(g,\eta,\xi,\Phi)$.  Then it follows that there exists
a smooth family of compatible Sasaki structures $(g_t,\eta_t,\xi,\Phi_t)\in\mathcal{S}(\xi,\bar{J}_t),\
t\in V\subset\mathcal{B}$, where $V$ is a neighborhood of zero in $\mathcal{B}$, if and only if
the deformation is of $(1,1)$-type restricted to $V$. \end{theorem}

Recall that an important class of deformations of a Sasaki structure $(g,\eta,\xi,\Phi)$ is given by the
$D$-homothetic transformations introduced by Tanno \cite{TAN}. A $D$-homothetic transformation, also
called $0$-type deformation, is defined for any positive constant $a$ as
\[
\bar{\eta}=a\eta,\ \bar{\xi}=\frac{1}{a}\xi,\ \bar{\Phi}=\Phi,\ \bar{g}=ag+a(a-1)\eta\otimes\eta.
\]

Other deformations of interest in Sasaki geometry are obtained deforming the Reeb vector field in the
Sasaki cone \cite{VCOV}. We recall that
those transformations that deform the characteristic foliation $\mathcal{F}_\xi$ are known as deformations
of type I  \cite{BGM}, they being firstly investigated by Takahashi \cite{TAK}. We note that in the
following we will focus on the deformations that preserve the Reeb vector field $\xi$, and consequently
the foliation $\mathcal{F}_\xi$. We are interested in deforming the transverse
K\"{a}hler structures by using a transverse K\"{a}hler-Ricci flow, called the Sasaki-Ricci flow.

\subsection{Sasaki-Ricci flow}

In analogy to the K\"{a}hler-Ricci flow \cite{CAO}, one can define a Sasaki-Ricci flow which preserves 
the Sasaki condition, in the sense that the evolved metrics remain Sasaki. Recall that this flow has 
been investigated in detail in \cite{S-W-Z}. 

Let $M$ be a $(2n+1)$-dimensional smooth manifold equipped with a Sasakian structure $(g,\eta,\xi,\Phi)$.
Suppose that we deform the contact form $\eta$ with a basic function $\varphi$ as follows:
\begin{equation}\label{etadef}
\tilde{\eta}=\eta +d_{B}^{c}\varphi.
\end{equation}

The above deformation implies that other fundamental tensors are also modified:
$$
\begin{array}{rcl}
\tilde{\Phi} & = & \Phi - (\xi \otimes (d_{B}^{c}\varphi)) \circ \Phi \, ,\\
\tilde{g} & =  & d\tilde{\eta}\circ (\BOne \otimes \tilde{\Phi} ) +
\tilde{\eta}\otimes \tilde{\eta} \, ,
\end{array}
$$
as well as the transverse form:
\[
d\tilde{\eta}=d\eta+d_{B}d_{B}^{c}\varphi.
\]

It is known (see, e.g., \cite{S-W-Z}) that the quadruplet $(\tilde{g},\tilde{\eta},\xi,\tilde{\Phi})$
remains a Sasakian structure on $M$.

Now, let $(g(t),\eta(t),\xi,\Phi(t))$ be a flow having initial data $(g(0),\eta(0),\xi,\Phi(0))=(g,\eta,\xi,\Phi)$,
generated by a basic function $\varphi(t)$ as above and suppose that the basic first Chern class is positive,
i.e. $c_B^1>0$.
Then the Sasaki-Ricci flow, also known as transverse
K\"{a}hler-Ricci flow, is defined by \cite{FOW}
\[
\frac{\partial g^T}{\partial t}=-Ric^T_{g(t)}+(2n+2)g^T(t),
\]
where $Ric^T$  is the transverse Ricci curvature.

Notice that, in terms of local coordinates, the above flow can be written as a transverse parabolic
Monge-Amp\`{e}re equation on the potentials.

\section{Deforming the orthogonal complement of the leafwise distribution on Sasaki-Einstein spaces}

In the following we investigate a method of deforming the metric $g$ on an open subset $U\in M$, by
keeping the transverse and the leafwise metric intact. We show that a
geometric description of such procedure is represented by the deformation of
the orthogonal complement of the leafwise distribution $L_{\xi} $, associated
to a particular basic function.

More precisely, let us consider a basic function $\varphi $ on $U$, which satisfies the equation
\begin{equation}
d_{B}d_{B}^{c}\varphi =\partial_B \bar{\partial}_B \varphi=0.  \label{harmonic_fct}
\end{equation}

In the sequel we denote by $\tilde{\mathcal{D}}$ the deformed complementary distribution on $U$. 
This specific deformation on a Riemannian foliation can be described in a standard way (see e.g. \cite{Al}). 

If we consider the vector $V \in \nu$, then we assigning to each corresponding vector 
$\sigma (V)\in \mathcal{D}$ (with $p(\sigma (V))=V$)
its deformed image $\tilde{\sigma}(V)\in \mathcal{\tilde{D}}$,
(with $\tilde{\sigma}:\nu \rightarrow \tilde{D}$ and $p(\tilde{\sigma}(V))=V$) in the following manner:
\begin{equation}
\tilde{\sigma}(V):=\sigma (V)-d_{B}^{c}\varphi (\sigma (V))\xi .
\label{variatie_comp_ortog}
\end{equation}
We show now that the associated deformed contact form $\tilde{\eta}$ will be given by \eqref{etadef}.
Indeed, we first notice that the morphism generated between vector spaces
$\mathcal{D}_{p}$ and $\mathcal{\tilde{D}}_{p}$ at each $p\in M$ is
injective, as we have the direct sum \eqref{splitting}.
Furthermore,
\[
\begin{split}
\tilde{\eta}(\tilde{\sigma}(V))& =(\eta +d_{B}^{c}\varphi )(\sigma
(V)-d_{B}^{c}\varphi (\sigma (V))\xi ) \\
&= \eta (\sigma (V))-d_{B}^{c}\varphi (\sigma (V))\eta (\xi ) \\
& +d_{B}^{c}\varphi (\sigma (V))-(d_{B}^{c}\varphi (\sigma
(V)))d_{B}^{c}\varphi (\xi ) \\
& =0,
\end{split}
\]
so $\mathrm{Ker}\,\eta ={\tilde{D}}$.

\begin{remark}
It is also possible to express the deformation \eqref{variatie_comp_ortog} using local
computation. Following the notations from \cite{S-W-Z}, if $U$ is a local chart with the 
corresponding local coordinates $(x,z^1,\ldots,z^n)$, with the corresponding local frame 
$(\frac{\partial}{\partial x},\frac{\partial}{\partial z^1},\ldots,\frac{\partial}{\partial z^n},
\frac{\partial}{\partial \bar{z}^1},\ldots,\frac{\partial}{\partial \bar{z}^n})$, then
the distribution ${\mathcal{D}^{\mathbb{C}}}$ is spanned by $X_j=\frac{\partial}{\partial z^j}+i K_j 
\frac{\partial}{\partial x}$. We get
\[
\begin{split}
-d_{B}^{c}\varphi \left( \frac{\partial }{\partial z^{j}}\right) & =-\frac{1}{2}i(\bar{\partial}_{B}-\partial _{B})\varphi 
\left( \frac{\partial }{\partial z^{j}}\right) \\
& =\frac{1}{2}i(\partial _{B}\varphi )\left( \frac{\partial }{\partial z^{j}}\right) \\
& =i\frac{1}{2}\sum_{k}\varphi _{k}dz^{k}\left(
\frac{\partial }{\partial z^{j}}\right) \\
& =i\frac{1}{2}\varphi _{j},
\end{split}
\]
and the deformed complementary distribution $\tilde{D^{\mathbb{C}}}$ is span by the complex local
vector fields
\[
\begin{split}
\tilde{X}_{j}& =\frac{\partial }{\partial z^{j}}
+i(K_{j}+\frac{1}{2}\varphi_{j})\frac{\partial}{\partial x} \\
& =X_{j}+ i\frac{1}{2}\varphi _{j}\frac{\partial}{\partial x} .
\end{split}
\]
\end{remark}

We also consider the deformed contact structure
\[
\tilde{\Phi}=\Phi -(\xi \otimes d_{B}^{c}\varphi) \circ \Phi .
\]

From \cite{B-G-S} we see that
\[
\tilde{\Phi}\circ \tilde{\eta}=0.
\]
Using \eqref{variatie_comp_ortog}, we can also easily check the invariance
of the complex structure \thinspace $J$ stated in \cite{S-W-Z} (see also \cite{Bel}).

We have
\[
\begin{split}
\tilde{\Phi}(\tilde{\sigma}(V)) &=(\Phi -(\xi \otimes d_{B}^{c}\varphi) \circ
\Phi ) \circ (\sigma (V)-d_{B}^{c}\varphi (\sigma (V))\xi ) \\
&=\Phi (\sigma (V))-(d_{B}^{c}\varphi \circ \Phi (\sigma (V)))\xi
-d_{B}^{c}\varphi (\sigma (V))\Phi (\xi ) \\
&+(d_{B}^{c}\varphi (\sigma (V)))(d_{B}^{c}\varphi \circ \Phi (\xi ))\xi \\
&=\Phi (\sigma (V))-(d_{B}^{c}\varphi \circ \Phi (\sigma (V)))\xi,
\end{split}
\]
because $\Phi (\xi )=0$. Then $p\circ \tilde{\Phi}\circ \tilde{\sigma}=p\circ \Phi \circ \sigma $, as $p(\xi )=0$, so
\[
JV=(\Phi \sigma (V))^{p}=(\tilde{\Phi}\tilde{\sigma}(V))^{p}.
\]
From \eqref{variatie_comp_ortog} we infer that
\[
\begin{split}
d\tilde{\eta} &=d\eta +dd_{B}^{c}\varphi \\
&=d\eta +d_{B}d_{B}^{c}\varphi \\
&=d\eta ,
\end{split}
\]
as $d_{B}^{c}\varphi $ is a basic 1-form and $d_{B}=d_{\mid \Omega _{B}}$ (see e.g. \cite{Al}).

The deformed metric is constructed in a standard manner using the contact form \cite{B-G-S}
\begin{equation} \label{deformed_metric}
\begin{split}
\tilde{g} &=d\tilde{\eta}(\mathrm{Id}\otimes \tilde{\Phi})+\tilde{\eta}
\otimes \tilde{\eta} \\
&=d\eta (\mathrm{Id}\otimes \tilde{\Phi})+\tilde{\eta}\otimes \tilde{\eta}.
\end{split}
\end{equation}

The transverse part of the metric $\tilde{g}^{T}$ can be expressed in
local coordinates as (see e.g. \cite{S-W-Z})

\[
\begin{split}
\tilde{g}^{T} &=\sum_{j,l=1}^{n}(g_{j,\bar{l}}^{T}+
\frac{\partial \varphi }{\partial z^{j}\partial \bar{z}^{l}})dz^{j}d\bar{z}^{l} \\
&=\sum_{j,l=1}^{n}g_{j,\bar{l}}^{T}dz^{j}d\bar{z}^{l}=g^{T},
\end{split}
\]
according again to \eqref{harmonic_fct}.

In the sequel we denote by $Ric_{g}$ the Ricci tensor on the Sasaki manifold
$M$ of dimension $2n+1$, and by $Ric^{T}$ the intrinsic Ricci tensor of the
transverse metric $g^{T}$.

The result below is a direct consequence of \cite[Proposition 2.3]{B-G-S}.

\begin{proposition} \label{Proposition_Boyer}
If $X$, $Y\in \mathcal{D}$, and $\xi $ is the Reeb vector field of the Sasaki
structure, then

i) $Ric_{g}(X,Y)=Ric^{T}(\sigma (x),\sigma (Y))-2g(X,Y)$,

ii)$Ric_{g}(X,\xi )=0$,

iii) $Ric_{g}(\xi ,\xi )=2n$.

\end{proposition}

\begin{remark}
As a conclusion, the Ricci tensor on a Sasaki manifold is completely
determined by the transverse metric $g^{T}$ and the transverse Ricci tensor $Ric^{T}$.
Then, if we deform the Sasaki structure in such a way that the
transverse metric is preserved (so $Ric^{T}$ is also preserved), then all
properties related to the $Ric_{g}$ (including the property of the metric $g$
to be a Sasaki-Einstein metric) will hold. We obtain the following result.
\end{remark}

\begin{proposition}
If the Sasaki-Einstein metric $g$ of a manifold is deformed on the open subset $U$ as in \eqref{deformed_metric},
then the new metric $\tilde{g}$ defined on $U$ will remain Sasaki-Einstein.
\end{proposition}

In the next sections we use the above results to construct families of
Sasaki-Einstein metrics on the classical  five-dimensional space $T^{1,1}$. Furthermore, we
will show that these families of metrics can be used to construct solutions
of the Sasaki-Ricci flow equation.

\section{Local coordinates on $T^{1,1}$}

We exemplify all the result on the classical Sasaki-Einstein manifold
represented by the five-dimensional space $T^{1,1}$. Consequently, 
all our further considerations and results will be local.
We recall that $T^{1,1}=S^2\times S^3$ is one of the most renowned example 
of homogeneous Sasaki-Einstein
space in dimension five, the standard metric on this manifold being \cite{CO,M-S}
\[
\begin{split}
ds^2(T^{1,1}) = & \frac16 (d \theta^2_1 + \sin^2 \theta_1 d \phi^2_1 +
d \theta^2_2 + \sin^2 \theta_2 d \phi^2_2) \\
& +\frac19 (d \psi + \cos \theta_1 d \phi_1 + \cos \theta_2 d \phi_2)^2
\,,
\end{split}
\]
where $\theta _{i}\in [0,\pi )$, $\phi _{i}\in [0,2\pi )$, $i=1,2$ and
$\psi \in [0,4\pi)$.

Following \cite{G-K-N}, we consider on $T^{1,1}$ a patch of
coordinates $(\psi $,$w^{1}$,$w^{2})$, where the real coordinates $\psi $ is
for the Reeb flow of the Sasaki structure, with $\frac{\partial }{\partial
\psi }=3\xi $, and $(w^{1}$, $w^{2})$ are transverse complex coordinates
addressing the transverse K\"{a}hler structure. As on $T^{1,1}$ the transverse
structure are locally isomorphic to a product $S^{2}\times S^{2}$, we choose
\begin{equation} \label{complex_coord}
\begin{split}
w^{1} &=\tan \frac{\theta _{1}}{2}e^{i\phi _{1}},  \\
w^{2} &=\tan \frac{\theta _{2}}{2}e^{i\phi _{2}}.
\end{split}
\end{equation}
We then get
\begin{eqnarray}
dw^{1} &=&\left( \frac{1}{2}\frac{1}{\cos ^{2}\frac{\theta _{1}}{2}}d\theta
_{1}+i\tan \frac{\theta _{1}}{2}d\phi _{1}\right) e^{i\phi _{1}}\text{,}
\label{d_omega} \\
dw^{2} &=&\left( \frac{1}{2}\frac{1}{\cos ^{2}\frac{\theta _{2}}{2}}d\theta
_{2}+i\tan \frac{\theta _{2}}{2}d\phi _{2}\right) e^{i\phi _{2}}\text{.}
\notag
\end{eqnarray}

From \eqref{complex_coord} we also get that
\begin{eqnarray*}
\frac{\partial }{\partial \theta _{j}} &=
&\frac{1}{2}\frac{1}{\cos ^{2}\frac{\theta _{j}}{2}}\left( \frac{\partial }{\partial w^{j}}e^{i\phi _{j}}
+\frac{\partial }{\partial \bar{w}^{j}}e^{-i\phi _{j}}\right) , \\
\frac{\partial }{\partial \phi _{j}} &=&\tan \frac{\theta _{j}}{2}\left( i\frac{\partial }
{\partial w^{j}}e^{i\phi _{j}}-i\frac{\partial }{\partial\bar{w}^{j}}e^{-i\phi _{j}}\right) ,
\end{eqnarray*}
for $1\leq j\leq 2$.

From here we derive the following local description of the derivatives $\frac{\partial }{\partial w^{j}}$,
$\frac{\partial }{\partial \bar{w}^{j}}$ which will be useful in our further computation.
\begin{equation}
\begin{split}
\frac{\partial }{\partial w^{j}}& =-i\cos ^{2}\frac{\theta _{j}}{2}
e^{-i\phi_{j}}\left( i\frac{\partial }{\partial \theta _{j}}+\frac{1}
{\sin \theta _{j}}\frac{\partial }{\partial \phi _{j}}\right) , \\
\frac{\partial }{\partial \bar{w}^{j}}& =-i\cos ^{2}\frac{\theta _{j}}{2}e^{i\phi _{j}}\left( i\frac{\partial }
{\partial \theta _{j}}-\frac{1}{\sin \theta _{j}}\frac{\partial }{\partial \phi _{j}}\right) .
\end{split}
\label{derivatives_w}
\end{equation}

Let us now compute the contact form $\eta $. We consider the Sasaki potential
\[
K= \frac13 \sum_{j} \log (1+w^{j}\bar{w}^{j})-\frac{1}{6}\sum_{j}\log(w^{j}\bar{w}^{j}).
\]
It gets, in accordance with \cite{G-K-N}
\[
\begin{split}
\eta  &=\frac{1}{3}d\psi +i\sum_{j}\frac{\partial K}{\partial w^{j}}dw^{j}
-i\sum_{j}\frac{\partial K}{\partial \bar{w}^{j}}d\bar{w}^{j} \\
&=\frac{1}{3}d\psi +i\frac{1}{3}\sum_{j}\frac{\bar{w}^{j}}{1+w^{j}\bar{w}^{j}}dw^{j}
-i\frac{1}{3}\sum_{j}\frac{w^{j}}{1+w^{j}\bar{w}^{j}}d\bar{w}^{j} \\
&-i\frac{1}{6}\sum_{j}\frac{1}{w^{j}}dw^{j}+i\frac{1}{6}\sum_{j}\frac{1}{\bar{w}^{j}}d\bar{w}^{j}.
\end{split}
\]

Using now \eqref{complex_coord} and \eqref{d_omega} we obtain
\[
\begin{split}
\eta  &=\frac{1}{3}d\psi
+\sum_{j}\left[ \left( \frac{1}{3}\cos ^{2}\frac{\theta _{j}}{2}\tan \frac{\theta _{j}}{2}
-\frac{1}{6}\frac{1}{\tan \frac{\theta _{j}}{2}}\right) e^{-i\phi _{j}}\right.  \\
&\left. \cdot \left( \frac{1}{2}\frac{1}{\cos ^{2}\frac{\theta _{j}}{2}}d\theta _{j}
+i\tan \frac{\theta _{j}}{2}d\phi _{1}\right) e^{i\phi _{j}}\right]  \\
&-\sum_{j}\left[ \left( \frac{1}{3}\cos ^{2}\frac{\theta _{j}}{2}\tan \frac{\theta _{j}}{2}
-\frac{1}{6}\frac{1}{\tan \frac{\theta _{j}}{2}}\right)e^{i\phi _{j}}\right.  \\
&\left. \cdot \left( \frac{1}{2}\frac{1}{\cos ^{2}\frac{\theta _{j}}{2}}d\theta _{j}
-i\tan \frac{\theta _{j}}{2}d\phi _{1}\right) e^{-i\phi _{j}}\right] .
\end{split}
\]
After computation, the contact form can be written%
\begin{equation} \label{eta_coord_real}
\begin{split}
\eta  &=\frac{1}{3}d\psi -\frac{2}{3}\sum_{j}\sin ^{2}\frac{\theta _{j}}{2}d\phi _{j}
+\frac{1}{3}\sum_{j}d\phi _{j} \\
&=\frac{1}{3}d\psi -\frac{1}{3}\sum_{j}\left( 1-2\sin ^{2}\frac{\theta _{j}}{2}\right) d\phi _{j}  \notag \\
&=\frac{1}{3}d\psi +\frac{1}{3}\sum_{j}\cos \theta _{j}\phi _{j}.
\end{split}
\end{equation}

Following the notations adopted in \cite{G-K-N}, the metric associated to
the above chosen Sasaki potential is
\begin{equation}\label{ds^2}
\begin{split}
ds_{T^{1,1}}^{2} &=\eta \otimes \eta +
2\sum_{j,\bar{l}}\frac{\partial ^{2}K}{\partial w^{j}\partial \bar{w}^{l}}dw^{j}d\bar{w}^{l} \\
&=\eta \otimes \eta +2\sum_{j}\frac{\partial ^{2}K}{\partial w^{j}\partial
\bar{w}^{j}}dw^{j}d\bar{w}^{j},
\end{split}
\end{equation}
As the first term is readily computable from \eqref{eta_coord_real}, we
express the second term.
\begin{equation} \nonumber
\begin{split}
\frac{\partial ^{2}K}{\partial w^{j}\partial \bar{w}^{j}} 
&=\frac13 \frac{1}{\left( 1+w^{j}\bar{w}^{j}\right) ^{2}} \\
&=\frac13 \cos ^{4}\frac{\theta _{j}}{2}.
\end{split}
\end{equation}
On the other hand, using \eqref{d_omega}, after computations we get
\[
\begin{split}
dw^{j}d\bar{w}^{j} &=\frac{1}{4}\frac{1}{\cos ^{4}\frac{\theta _{j}}{2}}d\theta _{j}^{2}
+\tan ^{2}\frac{\theta _{j}}{2}d\phi _{j}^{2} \\
&=\frac{1}{4}\frac{1}{\cos ^{4}\frac{\theta _{j}}{2}}\left( d\theta_{j}^{2}+\sin ^{2}\theta _{j}d\phi _{j}^{2}\right) .
\end{split}
\]
Then, plugging the above results back in \eqref{ds^2}, we obtain the
standard metric on $T^{1,1}$.

\begin{remark}
Alternatively, we may take the Sasaki potential to be
\[
K=\frac13 \sum_{j} \log (1+w^{j}\bar{w}^{j}),
\]
and choose to modify the leafwise constant accordingly, obtaining then a gauge transformation, as
suggested in \cite{G-K-N}. However, as throughout this paper we investigate
families of Sasaki metrics on $T^{1,1}$, we prefer to fix once for all the
local coordinates and choose always to modify the potential.
\end{remark}

\section{Families of Sasaki-Einstein metrics}

First of all, we investigate the general deformation of the orthogonal
complement to the leaves associated to a basic function $\varphi $ satisfying the equation \eqref{harmonic_fct}.
From \eqref{d_omega} and \eqref{derivatives_w} we get
\[
\begin{split}
\frac{\partial \varphi }{\partial w^{j}}dw^{j}
&=\frac{1}{2}\frac{\partial
\varphi }{\partial \theta _{j}}d\theta _{j}+\frac{\partial \varphi }
{\partial \phi _{j}}d\phi _{j}+\frac{1}{2}i\sin \theta _{j}\frac{\partial
\varphi }{\partial \theta _{j}}d\phi _{j} \\
&-i\frac{1}{\sin \theta _{j}}\frac{\partial \varphi }{\partial \phi _{j}}d\theta _{j} ,\\
\frac{\partial \varphi }{\partial \bar{w}^{j}}d\bar{w}^{j}
&=\frac{1}{2}\frac{\partial \varphi }{\partial \theta _{j}}d\theta _{j}
+\frac{\partial \varphi }{\partial \phi _{j}}d\phi _{j}
-\frac{1}{2}i\sin \theta _{j}\frac{\partial \varphi }{\partial \theta _{j}}d\phi _{j} \\
&+i\frac{1}{\sin \theta _{j}}\frac{\partial \varphi }{\partial \phi _{j}} d\theta _{j}.
\end{split}
\]

Using the above relations we compute the deformed contact form using the
formula \eqref{etadef} (see e.g. \cite{S-W-Z})
\[
\tilde{\eta}=\eta -i\frac12 \sum_{j}\frac{\partial \varphi }{\partial w^{j}}dw^{j}
+i\frac12 \sum_{j}\frac{\partial \varphi }{\partial \bar{w}^{j}}d\bar{w}^{j}.
\]
We obtain
\[
\begin{split}
\tilde{\eta} &=\eta +\frac12 \sum_{j}\sin \theta _{j}\frac{\partial \varphi }
{\partial \theta _{j}}d\phi _{j}-\sum_{j}\frac{1}{\sin \theta _{j}}\frac{\partial \varphi }
{\partial \phi _{j}}d\theta _{j} \\
&=\frac{1}{3}d\psi +\sum_{j}\left( \frac{1}{3}\cos \theta _{j}
+\frac12 \sin \theta_{j} \frac{\partial \varphi }{\partial \theta _{j}}\right)d\phi _{j} \\
&-\sum_{j}\frac{1}{\sin \theta _{j}}\frac{\partial \varphi }{\partial \phi
_{j}}d\theta _{j}.
\end{split}
\]
Consequently, we get the next result:

\begin{proposition}\label{Prop_basic_fct}
If $\varphi $ is a basic local function defined on the local complex chart considered above, 
satisfying the equation \eqref{harmonic_fct}, with respect to the complex coordinates given 
in the previous section, then any metric of the form
\[
\begin{split}
g &=\left( \frac{1}{3}d\psi +\sum_{j}\left( \frac{1}{3}\cos \theta _{j}
+\frac12 \sin \theta _{j} \frac{\partial \varphi }{\partial \theta _{j}}\right)d\phi _{j}\right.  \\
&\left. -\sum_{j}\frac{1}{\sin \theta _{j}}\frac{\partial \varphi }
{\partial \phi _{j}}d\theta _{j}\right) ^{2}+\frac{1}{6}\sum_{j}\left(d\theta _{j}^{2}
+\sin^2 \theta _{j}d\phi _{j}^{2}\right),
\end{split}
\]
defined on the local chart can be obtained by deforming the canonical metric structure on the manifold $T^{1,1}$.
Furthermore, in accordance with Proposition \ref{Proposition_Boyer}, it is Sasaki-Einstein.
\end{proposition}

In the following we discuss two convenient particular situations.

The first is suggested by the\ variation of the Sasaki potential in the
previous section. Namely, employing the notations from \cite{S-W-Z}, we take
the new potential $\tilde{K}$ to be
\[
\tilde{K}=K-\frac{1}{6}\sum_{j}c_{j}\log w^{j}\bar{w}^{j},
\]
where $c_{j}$ are arbitrary real constants, $j=1$, $2$. In the following, we
denote the additional Sasaki potential by $\varphi $. Then we compute now the
new contact form $\tilde{\eta}$, using the formula
\[
\begin{split}
\tilde{\eta} &=\eta -i\frac12 \sum_{j}\frac{\partial \varphi }{\partial w^{j}}dw^{j}
+i\frac12 \sum_{j}\frac{\partial \varphi }{\partial \bar{w}^{j}}d\bar{w}^{j} \\
&=\eta +i\frac{1}{12}\sum_{j}c_{j}\frac{dw^{j}}{w^{j}}
-i\frac{1}{12}\sum_{j}c_{j}\frac{d\bar{w}^{j}}{\bar{w}^{j}} \\
&=\eta +i\frac12 \sum_{j}c_{j}\frac{1}{\tan \frac{\theta _{j}}{2}}e^{-i\phi_{j}}dw^{j} \\
&-i\frac12 \sum_{j}c_{j}\frac{1}{\tan \frac{\theta _{j}}{2}}e^{i\phi _{j}}d\bar{w}^{j}.
\end{split}
\]
Introducing now \eqref{complex_coord} and \eqref{d_omega}, we obtain after
computations
\[
\begin{split}
\tilde{\eta}
&=\eta +\frac{1}{6}\sum_{j}c_{j}d\phi _{j} \\
&=\frac{1}{3}d\psi +\frac13 \sum_{j}\cos \theta _{j}d\phi _{j}
+\frac{1}{6}\sum_{j}c_{j}d\phi _{j}.
\end{split}
\]

Consequently, we get the following result.
\begin{proposition}
Any metric of the form
\[
\begin{split}
g &=\frac{1}{9}\left( d\psi +\sum_{j}(\cos \theta _{j}+\frac12 c_{j})d\phi_{j}\right) ^{2} \\
&+\frac{1}{6}\sum_{j}\left( d\theta _{j}^{2}+\sin^2 \theta _{j}d\phi^{2}_{j}\right),
\end{split}
\]
with arbitrary real constants $c_j$, with $j=1$, $2$, defined on the local chart considered above 
can be obtained by the deformation of the canonical metric on $T^{1,1}$. 
The deformed metric remains Sasaki-Einstein.
\end{proposition}

Next, we consider the additional potential $\varphi $ in the
following manner.
\[
\varphi =\sum_{j}c_{j}\log ^{2}w^{i}+\frac{1}{2}\sum_{j}c_{j}\log w^{j}\log
w^{j}\bar{w}^{j}-\frac{1}{4}\sum_{j}c_{j}\log ^{2}w^{j}\bar{w}^{j},
\]
with $c_{j}$ as above.

Let us notice that
\begin{equation} \label{part_phi}
\begin{split}
\frac{\partial \varphi }{\partial w^{j}}
&=-\sum_{j}c_{j}\frac{\log w^{j}}{w^{j}}
+\frac{1}{2}\sum_{j}c_{j}\left( \frac{1}{w^{i}}\log w^{j}\bar{w}^{j}+\frac{\log w^{j}}{w^{j}}\right)  \\
&-\frac{1}{2}\sum_{j}c_{j}\frac{\log w^{j}\bar{w}^{j}}{w^{j}}  \\
&=-\frac{1}{2}c_{j}\sum_{j}\frac{\log w^{j}}{w^{j}},
\end{split}
\end{equation}
and, similarly,
\begin{equation}\label{part_phi_conj}
\begin{split}
\frac{\partial \varphi }{\partial \bar{w}^{j}}
&=\frac{1}{2}\sum_{j}c_{j}\left( \frac{\log w^{i}}{\bar{w}^{i}}
-\frac{1}{2}\frac{\log w^{j}\bar{w}^{j}}{\bar{w}^{j}}\right)    \\
&=-\frac{1}{2}c_{j}\sum_{j}\frac{\log \bar{w}^{j}}{\bar{w}^{j}}.
\end{split}
\end{equation}

We get
\[
\begin{split}
\tilde{\eta} &=\eta -i\frac{1}{4}\sum_{j}c_{j}\frac{\log w^{j}}{w^{j}}dw^{j}
+i\frac{1}{4}\sum_{j}c_{j}\frac{\log \bar{w}^{j}}{\bar{w}^{j}}d\bar{w}^{j} \\
&=\eta -i\frac{1}{4}\sum_{j}c_{j}\left( \frac{e^{-i\phi _{j}}}{\tan \frac{\theta _{j}}{2}}\right)
(\log \tan \frac{\theta _{j}}{2}+i\phi _{j})dw^{j} \\
&+i\frac{1}{4}\sum_{j}c_{j}\left( \frac{e^{i\phi _{j}}}
{\tan \frac{\theta_{j}}{2}}\right) (\log \tan \frac{\theta _{j}}{2}-i\phi _{j})d\bar{w}^{j}.
\end{split}
\]

After plugging \eqref{d_omega} in the above relation, we end up with
\[
\begin{split}
\tilde{\eta}
&=\eta +\frac12 \sum_{j}c_{j}\frac{\phi _{j}}{\sin \theta _{j}}d\theta_{j}
+\frac12 \sum_{j}c_{j}\log \tan \frac{\theta _{j}}{2}d\phi _{j} \\
&=\frac{1}{3}d\psi +\frac13 \sum_{j}\cos \theta _{j}d\phi _{j} \\
&+\frac12 \sum_{j}c_{j}\frac{\phi _{j}}{\sin \theta _{j}}d\theta_{j}
+\frac12 \sum_{j}c_{j}\log \tan \frac{\theta _{j}}{2}d\phi _{j}.
\end{split}
\]

Obviously, from\eqref{part_phi} and \eqref{part_phi_conj} we see that $\varphi $ satisfies the
relation \eqref{harmonic_fct}. We proved the next statement.

\begin{proposition}
Any metric of the form
\[
\begin{split}
g &=\frac{1}{9}\left( d\psi +\sum_{j}\left( \cos \theta _{j}
+\frac12 c_{j}\log \tan \frac{\theta _{j}}{2}\right) d\phi _{j}
+\frac12 \sum_{j}c_{j}\frac{\phi _{j}}{\sin \theta _{j}}d\theta _{j}\right) ^{2} \\
&+\frac{1}{6}\sum_{j}\left( d\theta _{j}^{2} +\sin^2 \theta _{j}d\phi^{2} _{j}\right),
\end{split}
\]
with arbitrary real constants $c_j$, defined on the local chart constructed above is again Sasaki-Einstein.
\end{proposition}

\section{Transverse K\"{a}hler-Ricci flow}

It this section, as a final outcome, we show that the above deformations of
Sasaki-Einstein metrics can be also regarded as solutions of the equation of
the transverse K\"{a}hler-Ricci flow.

From \cite{S-W-Z}, this equation which involves the basic potential function $\varphi $ can be written 
in our particular setting as
\begin{equation} \label{Kahler_Ricci_flow_eq}
\frac{\partial \varphi }{\partial t}=\log \det (g_{j\bar{l}}^{T}
+\frac{\partial \varphi }{\partial w^{j}\partial \bar{w}^{l}})
-\log \det (g_{j\bar{l}}^{T})+6\varphi .
\end{equation}

We obtain the following result.

\begin{proposition}
On the Sasaki-Einstein manifold $T^{1,1}$, if the basic function $\varphi $ satisfies the condition 
\eqref{harmonic_fct}, then the flow $\varphi _{t}=(e^{6t}-1)\varphi $ satisfies equation
\eqref{Kahler_Ricci_flow_eq}.
\end{proposition}

The proof is straightforward, as for the particular case of the function $\varphi $  considered the equation becomes
\[
\frac{\partial \varphi }{\partial t}=6\varphi ,
\]
with the corresponding solution.

As a consequence, we have the following result.
\begin{corollary}
The families of potential basic functions
\[
\varphi _{t}=(e^{6t}-1)\sum_{j}c_{j}\log w^{j}\bar{w}^{j},
\]
and
\[
\varphi _{t}
=(e^{6t}-1)\Bigl[\sum_{j}c_{j}\log ^{2}w^{i}
+\frac{1}{2}\sum_{j}c_{j}\log w^{j}\log w^{j}\bar{w}^{j}
-\frac{1}{4}\sum_{j}c_{j}\log ^{2}w^{j}\bar{w}^{j} \Bigr] ,
\]
stand as solutions of the transverse K\"{a}hler-Ricci flow equation on the
manifold $T^{1,1}$.
\end{corollary}

\section{Conclusions}

In this paper we examine the K\"{a}hler structure of the transverse  K\"{a}hler geometry and
consider possible local deformations of the contact structure. We exemplify the general results in
the case of the five-dimensional Sasaki-Einstein space $T^{1,1}$. We introduce local holomorphic
coordinates and construct the Sasakian local potential, analogous to the K\"{a}hler potential.
We consider deformations of the contact form with a basic function. Choosing a basic function which 
satisfies \eqref{harmonic_fct}, we generate families of Sasaki-Einstein metrics of the form given in Proposition
\ref{Prop_basic_fct}. Moreover, two convenient particular situations are presented, giving the
expressions for the deformed local metrics. We remark that in the case of deformations with basic
functions as above we have an explicit solution of the equation of the Sasaki-Ricci flow.

In a forthcoming paper \cite{SVVprep} we shall consider deformations of the contact structures
modifying also the transverse part of the standard metric on $T^{1,1}$.

It is worth extending the study of deformations of the metric on the five-dimensional Sasaki-Einstein
spaces $Y^{p,q}$ as well as other contact structure as 3-Sasakian structures \cite{BG1999} or
 mixed 3-structures \cite{IVV}.

In general a system could possesses explicit and hidden symmetries encoded in the multitude of
Killing vectors and higher rank Killing tensors respectively. The complete sets of Killing-Yano
tensors were constructed on the five-dimensional Sasaki-Einstein spaces $T^{1,1}$ \cite{SVV2015}
and $Y^{p,q}$ \cite{MV,SVV2014,SVV2015AP}. It would be interesting to study the Killing forms on the
deformed contact structures and identify the corresponding hidden symmetries.

The deformations considered in this paper would be also interesting if they can be compared 
or perhaps connected with the so called $\beta$ or TsT (which consists of a T-duality, a
coordinate shift and another T-duality) deformations on Sasaki-Einstein manifolds \cite{DG}.
These deformations on Sasaki-Einstein spaces have important implications in holography and
string theory.

\section*{Acknowledgments}
VS and GEV were supported by CNCS-UEFISCDI,
project no. PN-III-P4-ID-PCE-2016-0065. MV is supported by the project NUCLEU PN 19 06 01 01/2019.

\end{document}